\documentclass[aps,prl,twocolumn,longbibliography]{revtex4-2}

\usepackage{amsmath,amssymb,bm}
\usepackage[colorlinks=true, pdfstartview=FitV, linkcolor=red, citecolor=blue, urlcolor=blue]{hyperref}
\usepackage{mathtools}
\usepackage[dvips]{graphicx}
\usepackage{hyperref}
\usepackage{yfonts}
\usepackage{color}
\usepackage{xcolor}
\usepackage{physics}
\usepackage{braket}
\usepackage{comment}
\newcommand\sect[1]{{\it #1}---}

\DeclareMathOperator{\e}{\mathrm{e}}
\DeclareMathOperator{\ii}{\mathrm{i}\!}

\newcommand{\beq}{\begin{eqnarray}}
\newcommand{\eeq}{\end{eqnarray}}
\renewcommand\d{\partial}

\begin{document}

\title{Quantum Metric and Nonlinear Hall Effect of Photons}

\author{Keidai Akiba and Naoki Yamamoto}
\affiliation{Department of Physics, Keio University, Yokohama 223-8522, Japan}

\begin{abstract}
Using the path-integral formalism, we show that photons possess a nontrivial quantum metric in momentum space. We derive the semiclassical action and equations of motion by taking into account the quantum metric.
In media with a spatially varying refractive index $n({\bm x})$, the quantum metric induces a shift in the trajectory of light at second order in spatial gradients of $n$, whose transverse component may be regarded as a nonlinear Hall effect of light. The quantum metric also gives rise to corrections to photon propagation relevant to gravitational lensing in curved spacetime at second order in wavelength, including a gravitational nonlinear Hall effect arising from the interplay between the geometry of position space and that of momentum space.
\end{abstract}
\maketitle

\sect{Introduction}%
Quantum geometry, characterized by the Berry curvature and quantum metric, is a cornerstone of modern condensed matter physics. The Berry curvature \cite{Berry} plays a key role in topological transport phenomena such as the anomalous Hall effect~\cite{Sundaram-Niu, Haldane:2004zz} and the chiral magnetic effect~\cite{Son:2012wh, Stephanov:2012ki}. 
On the other hand, the quantum metric \cite{Provost1980} governs nonlinear responses of quantum materials; see Refs.~\cite{Liu2024,Yu2025,Jiang2025,Verma2025,Gao2025} for recent reviews. 
Since not only electrons in solids but also photons possess nontrivial Berry curvature and exhibit anomalous Hall effect in inhomogeneous media \cite{Liberman:1992zz,Bliokh:2004gz,Onoda:2004zz,Duval:2005ky}, it is natural to ask whether photons also have a quantum metric and exhibit nonlinear transport associated with it.

In this paper, based on the path-integral formulation for the Hamiltonian of photons, we show that photons possess a nontrivial quantum metric in momentum space. 
While photons do not directly couple to electromagnetic fields unlike electrons in solids, their motion is affected by the spatially varying refractive index $n({\bm x})$ in medium, which leads to an optical analogue of the nonlinear Hall effect.
In addition, since a static weak background gravitational field can be regarded as a ``medium'' with an effective refractive index, the quantum metric also gives rise to corrections to the local photon dynamics underlying gravitational lensing at second order in wavelength. The transverse component of the local velocity correction constitutes a gravitational nonlinear Hall effect, resulting from the interplay between the geometry in position space and that in momentum space.

\sect{Hamiltonian of photons and 
path-integral formulation}%
We first recall the wave equation for the photon wave function \cite{Bialynicki-Birula,Yamamoto:2017uul}.
Generally, the wave equation for massless particles should satisfy the following requirements: (i) it must be linear so that the wave function can exhibit superposition and interference; (ii) its coefficients should be expressed in terms of physical constants, $\hbar$ and $c$, independent of the specific motion; (iii) it should reproduce the linear dispersion relation $\omega = c k$, where $\omega$ is the angular frequency and $k\coloneqq|{\bm k}|$ with $\bm k$ the wave vector. 

Corresponding to the right- and left-handed circular polarizations, ${\bm e}_{\pm} = {\bm e}_1 \pm \ii {\bm e}_2$, where ${\bm e}_{1,2}$ are the two polarization vectors of photons, the wave functions of right- and left-handed photons are taken to be proportional to ${\bm e}_{\pm}$. Accordingly, the wave functions of right- and left-handed one-photon states are three-component objects and are denoted by the vectors ${\bm \psi}_{\pm}$.
The photon wave equation satisfying all the above requirements can then be written in the form of Schr\"{o}dinger equation
$\ii \hbar \d_t {\bm \psi}_{\pm} = H_{\pm} {\bm \psi}_{\pm}$ \footnote{This photon wave equation is equivalent to the source-free Maxwell equations for the complex field combinations ${\bm E} \pm {\rm i}c\bm B$, corresponding to the helicities $\pm 1$~\cite{Bialynicki-Birula}.}, where 
\begin{equation}
\label{H_photon}
H_{\pm} = \pm c {\bm S} \cdot {\bm p}.
\end{equation}
Here, the subscripts $\pm$ denote the helicities $\pm 1$, ${\bm p}$ is the momentum, and $S_i$ ($i=1,2,3$) are the $3 \times 3$ matrices of the spin-$1$ representation, satisfying the commutation relations $[S_i, S_j] = \ii \epsilon_{ijk} S_k$. The physical photon states additionally satisfy the transversality condition ${\bm \nabla} \cdot {\bm \psi}_{\pm}=0$, which removes the longitudinal mode.

Let us then move on to the path integral formulation for the photon Hamiltonian~(\ref{H_photon})~\cite{Yamamoto:2017uul} (see also Ref.~\cite{Stephanov:2012ki} for the case of chiral fermions). In the following, we consider the Hamiltonian $H_+$ for right-handed photons as an example, but the discussion can be similarly applicable to $H_-$ for left-handed photons.
The transition amplitude from the initial state $i$ to the final state $f$ is written as
\begin{align}
    \label{eq:ta}
    \mathcal{A}&=\bra{f} \e^{-\frac{\ii}{\hbar} H(t_{\rm f}-t_{\rm i})} \ket{i} \notag \\
    &= \int \mathcal{D}\bm{x} \mathcal{D}\bm{p}\,{\cal P} \exp [\frac{\ii}{\hbar} \int^{t_{\rm f}}_{t_{\rm i}} \dd{t} (\bm{p} \cdot \bm{\dot{x}} - c \bm{S} \cdot \bm{p})]\,,
\end{align}
where ${\cal P}$ denotes the path-ordered product of the matrices $\bm S$.
The Hamiltonian $c \bm{S} \cdot \bm{p}$ can be diagonalized by introducing a unitary matrix $V_{\bm{p}}$ such that $V_{\bm{p}}^{\dagger} \bm{S} \cdot \bm{p} V_{\bm{p}} = |\bm{p}| \lambda_3$,
where $\lambda_3 = \textrm{diag}(1, -1, 0)$.
By decomposing the integral over $t$ into a sum over infinitesimal time intervals $\Delta t$ and inserting $1=V_{\bm{p}} V_{\bm{p}}^{\dagger}$ between successive exponential factors, we can rewrite
\begin{align}
    \label{eq:sum}
    &\cdots \e^{-\frac{\ii}{\hbar} \Delta t (c \bm{S}\cdot\bm{p})} \e^{-\frac{\ii}{\hbar} \Delta t (c \bm{S}\cdot\bm{p}')} \cdots 
    \nonumber \\
    &= \cdots V_{\bm{p}} \e^{-\frac{\ii}{\hbar} \Delta t (c|\bm{p}| \lambda_{3})} V_{\bm{p}}^{\dagger} V_{\bm{p}'} \e^{-\frac{\ii}{\hbar} \Delta t (c|\bm{p}'|\lambda_3)} V_{\bm{p}'}^{\dagger} \cdots
\end{align}
where $\Delta \bm{p} = \bm{p} - \bm{p}'$ is taken to be sufficiently small. 
Furthermore, by using
\begin{align}
    V_{\bm{p}}^{\dagger} V_{\bm{p}'} \simeq 1-V_{\bm{p}} {\bm \nabla}_{\bm{p}} V_{\bm{p}} \cdot \Delta{\bm{p}} \notag 
    \simeq \e^{-\ii \hat{\bm{a}}_{\bm{p}} \cdot \Delta \bm{p}},
\end{align}
with $\hat{\bm{a}}_{\bm{p}} = - \ii V_{\bm{p}}^{\dagger} {\bm \nabla}_{\bm{p}} V_{\bm{p}}$ being the connection in momentum space, we obtain 
\begin{gather}
    \mathcal{A} = V_{\bm{p}_{\rm f}} \int \mathcal{D}\bm{x} \mathcal{D}\bm{p}\,{\cal P} \exp \qty(\frac{\ii}{\hbar} \int^{t_{\rm f}}_{t_{\rm i}} \dd{t} L^{(1)}) V_{\bm{p}_{\rm i}}\,, \nonumber \\
L^{(1)} = \bm{p} \cdot \bm{\dot{x}} - c|\bm{p}|\lambda_{3} - \hbar \hat{\bm{a}}_{\bm{p}} \cdot \dot{\bm{p}}.
    \label{eq:ta1}
\end{gather}

\sect{Path-integral formulation at ${\cal O}(\hbar^{2})$ and quantum metric of photons}%
Based on the action~(\ref{eq:ta1}) and following the procedure for chiral fermions in Ref.~\cite{Mameda:2025rfn}, let us derive the semiclassical action at ${\cal O}(\hbar^2)$.
The connection $\hat{\bm{a}}_{\bm{p}}$ introduced above generally has off-diagonal components.
To transform the off-diagonal components at $\order{\hbar}$ into the diagonal components at $\order{\hbar^2}$, we perform a perturbative diagonalization due to Luttinger--Kohn \cite{Luttinger-Kohn} and Schrieffer--Wolff \cite{Schrieffer-Wolff}.
We first divide the Hamiltonian~\eqref{eq:ta1} as $H = H_{0} + \bar{H}_{1} + \delta H_1$,
where 
\begin{equation}
    H_0 = c|\bm{p}|\lambda_3, \quad \bar{H}_1 = \hbar \bar{\bm{a}}\cdot \dot{\bm{p}}, \quad \delta H_1 = \hbar \delta \bm{a} \cdot \dot{\bm{p}},
\end{equation}
with $\bar{\bm{a}}$ the diagonal components of $\hat{\bm{a}}$ defined as
$\bar{\bm{a}} = \bm{a} \lambda_3$, $\bm{a} = [\hat{\bm{a}}]_{11} = -[\hat{\bm{a}}]_{22}$ and $\delta \bm{a}$ the off-diagonal components.

We then introduce a unitary operator $\e^{S}$ with $S$ the anti-Hermitian matrix satisfying the relation
\begin{equation}
    \label{eq:Sdef}
    [S, H_{0}] + \delta H_{1} = 0.
\end{equation}
As in the calculation of Eq.~\eqref{eq:sum}, by inserting $1 = \e^{-S}\e^{S}$ between the exponential factors and using
\begin{equation}
    \e^{S_{\bm p}} \e^{-S_{\bm{p}'}} \simeq \e^{S_{\bm p}} \e^{- (S_{\bm p} - \Delta{\bm p} \cdot {\bm \nabla}_{\! \bm p} S_{\bm p})} \simeq \e^{\Delta \bm p \cdot {\bm \nabla}_{\! \bm p} S_{\bm p}} , \nonumber
\end{equation}
and 
\begin{align}
    \e^{S} \e^{-\frac{\ii}{\hbar} H \Delta t} \e^{-S} 
    = \e^{-\frac{\ii}{\hbar}\Delta t \qty[H_0 + \bar{H}_1 + [S, \bar{H}_1] + \frac{1}{2} [S, \delta H_1] +\order{\hbar^3} ]},
    \nonumber
\end{align}
we obtain
\begin{align}
    \mathcal{A} & = U_{{\bm p}_{\rm f}} \! \int \mathcal{D}\bm{x}\mathcal{D}\bm{p}\,{\cal P} \exp \qty(\frac{\ii}{\hbar}\int^{t_{\rm f}}_{t_{\rm i}}\dd{t} L^{(2)}) U_{{\bm p}_{\rm i}}, \nonumber \\
    L^{(2)} &= \bm{p} \cdot \dot{\bm x} - H_0 - \bar{H}_1 - [S,\bar{H}_1] - \frac{1}{2}[S, \delta H_1] \nonumber \\
    &\quad - \ii \hbar \dot{\bm{p}} \cdot {\bm \nabla}_{\! \bm p} S_{\bm p},
    \label{eq:NCKT-Lag1}
\end{align}
where $U = V \e^{-S}$.

We introduce $\ket{+1}, \, \ket{-1}$, and $\ket{0}$ as the eigenstates of the Hamiltonian $H_0$ with the eigenvalues $\pm c|\bm p|$ and $0$, respectively. For right-handed photons, only $\ket{+1}$ is the physical positive-energy photon state; $\ket{-1}$ and $\ket{0}$ correspond to the negative-energy and longitudinal sectors, respectively. They are nevertheless retained in the full three-component basis used for the perturbative diagonalization, because momentum derivatives of $\ket{+1}$ can have components outside the physical subspace. The effective action is eventually projected onto $\ket{+1}$.

We consider the matrix elements of the anti-Hermitian matrix $S$, defined by $S_{s,s'} = \bra{s} S \ket{s'}$ for $s,s'=\pm1, 0$.
Since the condition~\eqref{eq:Sdef} is satisfied for arbitrary values of the diagonal components of $S$, the diagonal matrix elements are not fixed. We therefore set $S_{s,s}=0$.
By contrast, the off-diagonal components of $S$ can be obtained from the condition~\eqref{eq:Sdef}.
We set $\epsilon_{s}=s c|\bm p| $ as the eigenvalues of $H_0$. Since $(\epsilon_{s'}-\epsilon_{s})S_{s,s'}+[\delta H_{1}]_{s,s'}=0$ for $s\neq s'$, we get 
\begin{equation}
    S_{s,s'} = \frac{[\delta H_1]_{s,s'}}{\epsilon_{s}-\epsilon_{s'}} = \frac{\hbar}{s-s'}\frac{\dot{\bm p} \cdot \hat{\bm a}_{s,s'}}{c|\bm p|}\,.
\end{equation}
At the end, we arrive at the following Lagrangian including the $\order{\hbar^2}$ corrections:
\begin{equation}
    \label{eq:NCKT-Lag2}
    L ^{(2)}= \bm{p} \cdot \dot{\bm x} - c|\bm p| \lambda_{3} - \hbar \bar{\bm a} \cdot \dot{\bm p} - \frac{\hbar^2}{2} \dot{p}_{i}\dot{p}_{j}\hat{G}_{ij}\,,
\end{equation}
where
\begin{widetext}
\begin{align}
    & \hat{G}_{ij} = \notag \\
    & \mqty(
    \frac{[\hat{a}_{(i}]_{+-} [\hat{a}_{j)}]_{-+} + 2 [\hat{a}_{(i}]_{+0} [\hat{a}_{j)}]_{0+}}{c|\bm p|} & (\ii \partial^{p}_{(i} - 2a_{(i}) \frac{[\hat{a}_{j)}]_{+-}}{c|\bm p|} & (2\ii \partial^{p}_{(i} - 2a_{(i}) \frac{[\hat{a}_{j)}]_{+0}}{c|\bm p|} + \frac{3}{2} \frac{[\hat{a}_{(i}]_{+0} [\hat{a}_{j)}]_{-0}}{c|\bm p|} \\ (-\ii \partial^{p}_{(i} - 2a_{(i}) \frac{[\hat{a}_{j)}]_{-+}}{c|\bm p|} & - \frac{[\hat{a}_{(i}]_{-+} [\hat{a}_{j)}]_{+-} + 2 [\hat{a}_{(i}]_{-0} [\hat{a}_{j)}]_{0-}}{c|\bm p|} & (-2\ii \partial^{p}_{(i} - 2a_{(i}) \frac{[\hat{a}_{j)}]_{-0}}{c|\bm p|} - \frac{3}{2} \frac{[\hat{a}_{(i}]_{-+} [\hat{a}_{j)}]_{+0}}{c|\bm p|} \\ (-2\ii \partial^{p}_{(i} - 2a_{(i}) \frac{[\hat{a}_{j)}]_{0+}}{c|\bm p|} + \frac{3}{2} \frac{[\hat{a}_{(i}]_{0-} [\hat{a}_{j)}]_{-+}}{c|\bm p|} & (2\ii \partial^{p}_{(i} - 2a_{(i}) \frac{[\hat{a}_{j)}]_{0-}}{c|\bm p|} - \frac{3}{2} \frac{[\hat{a}_{(i}]_{0+} [\hat{a}_{j)}]_{+-}}{c|\bm p|} & - \frac{2[\hat{a}_{(i}]_{0+} [\hat{a}_{j)}]_{+0} - 2 [\hat{a}_{(i}]_{0-} [\hat{a}_{j)}]_{-0}}{c|\bm p|}
    )
\end{align}
\end{widetext}
with $\d^p_i=\d/\d p_i$ and $X_{(i} Y_{j)} = \frac{1}{2}(X_{i}Y_{j}+X_{j}Y_{i})$.

The Lagrangian derived here still has off-diagonal components. However, by repeating the same argument, it can be shown that the $\order{\hbar^2}$ off-diagonal components do not contribute to the $\order{\hbar^2}$ diagonal components \cite{Mameda:2025rfn}. 
Focusing on the positive-energy state $\ket{+1}$, the $\order{\hbar^2}$ effective action for right-handed photons reads
\begin{equation}
    \label{eq:NCKT-action}
    I = \int^{t_{\rm f}}_{t_{\rm i}} \dd{t} \qty(\bm{p} \cdot \dot{\bm x} - c|\bm{p}| - \hbar \bm{a} \cdot \dot{\bm p} - \frac{\hbar^2}{2} \dot{p}_i \dot{p}_j G_{ij})\,,
\end{equation}
where
\begin{equation}
    G_{ij} \coloneqq \langle +|\hat{G}_{ij}|+ \rangle = 
    2\,\mathrm{Re}\sum_{{s'}\neq +} \frac{\langle \partial^p_i +|{s'}\rangle
\langle {s'}|\partial^p_j +\rangle}{\epsilon_{+}-\epsilon_{s'}}
\end{equation}
is the energy-normalized quantum metric.
This adiabatic projection is justified for ${\hbar}|\dot {\bm p}| \ll {c}|{\bm p}|^2$ \cite{Yamamoto:2017uul}. We can compute $G_{ij}$ by substituting the explicit form of $\hat{\bm{a}}$ as
\begin{align}
    \label{eq:QM}
    G_{ij} &= [\hat{G}_{ij}]_{++} = \frac{1}{c|\bm p|^3}(\delta_{ij} - \hat{p}_{i} \hat{p}_{j})\,,
\end{align}
where $\hat {\bm p} = {\bm p}/|{\bm p}|$ is the unit vector. This differs from the conventional quantum metric $g_{ij}\coloneqq \mathrm{Re}\sum_{{s'}\neq +} \langle \partial^p_i +|{s'}\rangle
\langle {s'}|\partial^p_j +\rangle$ \cite{Provost1980} by the inverse-energy-gap weighting of the intermediate states. For the helicity-1 photon, however, a single momentum derivative couples only adjacent spin-1 helicity sectors, so that $\langle -|\partial^p_i+\rangle=0$. Hence, only the state $|0\rangle$ contributes, and the two metrics are simply related by $G_{ij} = 2g_{ij}/(c|{\bm p}|)$.

We can similarly derive $G_{ij}$ for left-handed photons and find that it is also given by Eq.~(\ref{eq:QM}). Unlike the Berry curvature of photons, $\bm{\Omega} \coloneqq \bm{\nabla}_{\bm p} \times \bm{a} = \pm \hat {\bm p} / |{\bm p}|^2$, which has opposite signs for right- and left-handed photons, the quantum metric is the same for both helicities.
Note also that $G_{ij}$ for photons is four times larger than that for chiral fermions. 
This factor of four arises from two factors of two: the conventional quantum metric is twice as large for helicity-1 photons as for helicity-$1/2$ fermions, while the relevant energy gap is $c|{\bm p}|$ for photons and $2c|{\bm p}|$ for chiral fermions.

\sect{Nonlinear Hall effect of light in inhomogeneous media}%
Let us consider an isotropic inhomogeneous medium with spatially varying permittivity $\varepsilon({\bm x})$ and permeability $\mu({\bm x})$, for which the local propagation speed is $v({\bm x})=1/\sqrt{\varepsilon({\bm x})\mu({\bm x})}$, the refractive index is $n({\bm x})=c/v({\bm x})$, and the impedance is $Z({\bm x})=\sqrt{\mu({\bm x})/\varepsilon({\bm x})}$. Since spatial variations of $Z$ generally couple the two helicity sectors~\cite{Bialynicki-Birula}, we restrict the following analysis to constant-impedance media, ${\bm \nabla}Z=0$, while allowing $n({\bm x})$ to vary. The two helicity sectors then decouple. With the midpoint prescription for the phase-space path integral, the corresponding phase-space Hamiltonian is $H_{\pm} = \pm v({\bm x}){\bm S}\cdot{\bm p}$. Accordingly, the effective single-helicity action takes the form of Eq.~(\ref{eq:NCKT-action}) with $c$ replaced by $v({\bm x})$.

The local dispersion relation becomes $\epsilon_{\bm p} =v p$ with $p\coloneqq|{\bm p}|$.
The Hamilton equations are
\begin{align}
\label{eq:ELx}
\dot p_k &=-p\,\d_kv
-\frac{\hbar^2}{2}\dot p_i\dot p_j\d_kG_{ij},
\\
\label{eq:ELp}
\dot x_k &=v\hat p_k+\hbar\epsilon_{kij}\dot p_i\Omega_j \nonumber \\
& \quad -\hbar^2\left(
\dot p_i\dot p_j\Gamma_{ijk}
+\ddot p_jG_{kj}
+\dot x_i\dot p_j\d_i G_{kj}
\right),
\end{align}
where $G_{ij}=G_{ij}({\bm x}, {\bm p})$ and
\begin{align}
    \Gamma_{ijk} &\coloneqq \frac{1}{2} (\partial^{p}_{i}G_{jk} + \partial^{p}_{j}G_{ik} - \partial^{p}_{k}G_{ij}) 
    \nonumber \\ 
    &= \frac{\delta_{ij} \hat{p}_{k} - 3 \delta_{jk} \hat{p}_{i} - 3 \delta_{ik} \hat{p}_{j} + 5 \hat{p}_{i} \hat{p}_{j} \hat{p}_{k}}{2 v p^4}\,
        \label{eq:CS}
\end{align}
is the Christoffel symbol associated with $G_{ij}$ in momentum space.

Since $\d_k G_{ij}=-(\d_k v)G_{ij}/v$, the ${\cal O}(\hbar^2)$ correction in Eq.~\eqref{eq:ELx} is third order in spatial gradients and does not enter the velocity through second order in gradients. Using the leading equations $\dot p_k=-p\,\d_kv$ and $\dot x_k=v\hat p_k$, Eq.~\eqref{eq:ELp} becomes
\begin{align}
\label{eq:NCKT-eom}
\dot x_k &=v\hat p_k-\hbar p\epsilon_{ijk}(\d_iv)\Omega_j
+\hbar^2v(\d_i\d_jv)p_iG_{jk}
\nonumber\\
&\quad-\hbar^2(\d_iv)(\d_jv)
\left(p^2\Gamma_{ijk}+2p_iG_{jk}\right).
\end{align}
The second term at ${\cal O}(\hbar)$ is the optical Magnus effect \cite{Liberman:1992zz}, also known as the spin Hall effect of light \cite{Bliokh:2004gz, Onoda:2004zz, Duval:2005ky}. 

When ${\bm \nabla} n$ is approximately constant, $\d_i\d_jn\simeq0$, one has $\d_i\d_jv=({2}/{v})(\d_iv)(\d_jv)$. The term including $\d_i\d_jv$ in Eq.~\eqref{eq:NCKT-eom} then cancels the term proportional to $(\d_iv)(\d_jv)p_iG_{jk}$, and the ${\cal O}(\hbar^2)$ correction reduces to
\begin{equation}
\label{eq:medium-corrected}
\dot{\bm x}^{(2)}
=-\frac{\hbar^2}{2vp^2}
\left\{
\left[({\bm\nabla}v)^2
+5({\bm\nabla}v\cdot\hat{\bm p})^2\right]\hat{\bm p}
-6({\bm\nabla}v\cdot\hat{\bm p}){\bm\nabla}v
\right\}.
\end{equation}
Its transverse component with respect to ${\bm \nabla} v$ can be regarded as a nonlinear Hall effect induced by the inhomogeneous medium. 
Note that this nonlinear Hall response does not require intrinsic breaking of either parity or time-reversal symmetry of the medium.
In contrast to the conventional spin Hall effect of light, which is helicity odd and produces a shift perpendicular to both ${\bm \nabla} v$ and $\hat{\boldsymbol p}$, the quantum-metric contribution is helicity even and lies in the plane spanned by these two vectors. The ${\cal O}(\hbar)$ and ${\cal O}(\hbar^2)$ transverse corrections are therefore orthogonal to each other.

The quantum-metric correction also exhibits the characteristic scaling of the displacement over an effective propagation length $D$,
\begin{equation}\label{scaling}
\delta x_{\rm QM} \sim
D\left(\frac{\lambda}{2\pi L_n}\right)^2\,, 
\end{equation}
up to an angular factor of order unity. 
Here, $\lambda$ is the wavelength of light and $L_n=|{\bm \nabla}\ln n|^{-1}$ is the typical length scale of the refractive-index variation. This quadratic wavelength dependence provides an additional distinction from the conventional spin Hall shift, which is linear in the wavelength at leading order.

Helicity reversal, simultaneous measurements of the in-plane and out-of-plane beam-centroid shifts, and the scaling in Eq.~(\ref{scaling}) may provide complementary tests of the effect~\footnote{Finite-width wave packets can in general exhibit additional multipole corrections, including contributions from real-space quadrupole moments at ${\cal O}(\lambda)$, which depend on the beam profile~\cite{Collingbourne:2026pok}. Such contributions are physically distinct from the intrinsic quantum-metric correction obtained here, which does not involve quadrupole moments and is ${\cal O}(\lambda^2)$. Their profile dependence may be tested experimentally by varying the beam waist, ellipticity, and transverse orientation, providing a way to separate them from the intrinsic contribution.}.
For example, taking $\lambda\sim10\,\mu{\rm m}$, $D\sim100\,\mu{\rm m}$, and a relative refractive-index change $\Delta n/n \sim 0.05$ across this region, corresponding to $L_n\sim2\,{\rm mm}$, we obtain $\delta x_{\rm QM} \sim 6 \times 10^{-2}\,\mathrm{nm}$. The estimated displacement may be accessible with advanced optical metrology, although a quantitative sensitivity estimate at this wavelength requires an appropriate measurement strategy for the helicity-even shift.

\sect{Gravitational nonlinear Hall effect in a weak gravitational field}%
Based on our formalism above, we can also consider $\order{\hbar^2}$ finite-wavelength corrections to photon propagation in a gravitational field. We work in the geometric-optics regime, where the wavelength is much smaller than the curvature radius of the background spacetime. To treat the nonlinear gravitational response consistently through second order in the weak-field expansion, we consider the Schwarzschild geometry in isotropic coordinates. Writing $r=|{\bm x}|$ for the radial coordinate and $R_{\rm s}=2GM/c^2$, with $M$ the black-hole mass and $G$ the gravitational constant, the metric is
\begin{equation}
\dd s^2
=
\left(
\frac{1-\frac{R_{\rm s}}{4r}}
     {1+\frac{R_{\rm s}}{4r}}
\right)^2
c^2\dd t^2
-
\left(
1+\frac{R_{\rm s}}{4r}
\right)^4
(\dd{\bm x})^2 .
\end{equation}

For null propagation, this isotropic metric can be represented by a scalar optical refractive index~\cite{Carroll}%
\footnote{More generally, null propagation in a static spacetime is described by an optical spatial metric~\cite{Perlick:2004tq}. 
The scalar refractive-index representation used here applies to the isotropic Schwarzschild geometry.}. 
Introducing the Newtonian potential $\phi(r)=-GM/r$, the weak-field regime corresponds to $|\phi|/c^2=R_{\rm s}/(2r)\ll1$. The refractive index is then
\begin{equation}
n
=
\frac{\left(1+\frac{R_{\rm s}}{4r}\right)^3}
     {1-\frac{R_{\rm s}}{4r}}
=
1-\frac{2\phi}{c^2}
+\frac{7}{4}\frac{\phi^2}{c^4}
+{\cal O}\left(\frac{\phi^3}{c^6}\right),
\end{equation}
and the corresponding coordinate velocity $v\coloneqq |{\rm d}{\bm x}|/\dd t$ is
\begin{equation}
v
= \frac{c}{n}
= c\left[
1+\frac{2\phi}{c^2}
+\frac{9}{4}\frac{\phi^2}{c^4}
+{\cal O}\left(\frac{\phi^3}{c^6}\right)
\right].
\label{eq:grav-v2}
\end{equation}

Using the corrected medium equation of motion, the quantum-metric contribution can be written as
\begin{align}
\dot x_k^{(2)}
&=\frac{\hbar^2}{p^2}(\delta_{kj}-\hat p_k\hat p_j)(\d_i\d_jv)\hat p_i
\nonumber\\
&\quad-\frac{\hbar^2}{2vp^2}\left\{
\left[({\bm\nabla}v)^2+(\hat{\bm p}\cdot{\bm\nabla}v)^2\right]\hat p_k
-2(\hat{\bm p}\cdot{\bm\nabla}v)\d_kv
\right\}.
\label{eq:grav-qm-general}
\end{align}
Substituting Eq.~\eqref{eq:grav-v2} separates this correction into terms linear and quadratic in $\phi$. The term linear in $\phi$ is
\begin{equation}
\dot{\bm x}^{(2)}_{\phi}
=
\frac{6\hbar^2GM}{cp^2r^3}\,
\chi(\chi\hat{\bm p}-\hat{\bm r}), \quad\chi\coloneqq\hat{\bm p}\cdot\hat{\bm r}
\label{eq:grav-linear}
\end{equation}
while the term quadratic in $\phi$ is
\begin{equation}
\dot{\bm x}^{(2)}_{\phi^2}
=
-\frac{2\hbar^2}{c^3p^2}
\left(\frac{GM}{r^2}\right)^2
\left[(1+10\chi^2)\hat{\bm p}-11\chi\hat{\bm r}\right].
\label{eq:grav-nonlinear}
\end{equation}
The component of Eq.~\eqref{eq:grav-nonlinear} transverse to ${\bm\nabla}\phi\parallel\hat{\bm r}$ may be regarded as the gravitational nonlinear Hall effect due to the quantum metric. This correction is distinct from the previously known gravitational spin Hall effect of light at $\order{\hbar}$~\cite{Gosselin:2006wp,Yamamoto:2017gla,Oancea:2020khc}. 

The quantum-metric contributions linear and quadratic in $\phi$ have the characteristic magnitudes,
\begin{align}
\frac{{\cal M}^{(2)}_{\phi}}{{\cal M}^{(0)}}
\sim \!
\left(\frac{R_{\rm s}}{r}\right)^{\! \! 3}
\! \! \left(\frac{\lambda}{2\pi R_{\rm s}}\right)^{\! \! 2}, \quad
\frac{{\cal M}^{(2)}_{\phi^2}}{{\cal M}^{(0)}}
\sim \!
\left(\frac{R_{\rm s}}{r}\right)^{\! \! 4}\! \! \left(\frac{\lambda}{2\pi R_{\rm s}}\right)^{\! \! 2},
\label{eq:M2}
\end{align}
respectively. Here, ${\cal M}^{(n)}$ denotes the characteristic magnitude of the $\order{\hbar^n}$ contribution to $\dot{\bm x}$, and ${\cal M}^{(0)}=c/n$.
For example, as an order-of-magnitude estimate, for light with wavelength $\lambda\sim300\,{\rm m}$ passing at $r\sim5R_{\rm s}$ around a solar-mass Schwarzschild black hole, the two relative magnitudes are ${\cal M}^{(2)}_{\phi}/{\cal M}^{(0)}\sim10^{-6}$ and ${\cal M}^{(2)}_{\phi^2}/{\cal M}^{(0)}\sim10^{-7}$, respectively. 
Although both corrections are extremely small, their characteristic quadratic wavelength dependence distinguishes them from the geometric-optics contribution and provides a well-defined finite-wavelength signature that is relatively enhanced at longer wavelengths and in stronger gravitational fields.

\sect{Conclusion and outlook}%
In this paper, we derived the quantum metric for photons in momentum space and showed that it gives rise to a nonlinear Hall effect in inhomogeneous media and a gravitational nonlinear Hall effect in a weak gravitational field.
Unlike the linear Hall effect of light, which leads to splitting between right- and left-handed circular polarizations, this effect does not distinguish between the two.
For Schwarzschild spacetime, the $\order{\hbar^2}$ quantum-metric correction separates into terms linear and quadratic in the gravitational potential, with the transverse component of the latter identified as the gravitational nonlinear Hall effect. Relating these local velocity corrections to an observable lensing signal requires specifying the propagation geometry and integrating the equations of motion along the trajectory.

The $\order{\hbar}$ Berry-curvature corrections to the classical equations of motion have been derived from Maxwell's equations \cite{Onoda:2006ne,Nishida:2026bzu} and quantum field theory (quantum electrodynamics) \cite{Mameda:2022ojk} (see also Refs.~\cite{Huang:2020kik,Hattori:2020gqh}). It should be possible to extend such analyses to include the $\order{\hbar^2}$ corrections associated with the quantum metric.
As in the case of the Berry curvature of photons \cite{Onoda:2006ne}, the quantum-metric-induced correction may admit a formulation as a higher-order finite-wavelength effect purely within classical Maxwell's equations. 
Indeed, once the physical quantity is expressed in terms of the wavelength $\lambda$, as in Eq.~(\ref{eq:M2}), $\hbar$ no longer appears explicitly.
A direct derivation of the same quantum-metric correction from Maxwell’s equations remains to be established.

The present derivation of the quantum metric and the resulting nonlinear Hall effect for helicity-1 photons may be generalized to helicity-2 gravitational waves and, more broadly, to massless particles of arbitrary helicity, building on the approaches developed in Refs.~\cite{Yamamoto:2017gla,Huang:2018aly}.
We leave these issues for future work.

\sect{Acknowledgments}%
The author N.~Y. thanks Kazuya Mameda for collaboration on related works. 
This work was supported by JSPS KAKENHI Grant No.~JP24K00631.

\bibliography{photon}

\end{document}